\documentclass[letterpaper, 10 pt, conference]{ieeeconf} 
\usepackage{cite}
\usepackage{amsmath,amssymb,amsfonts}
\usepackage{algorithmic}
\usepackage{xcolor}
\usepackage{graphicx}

\definecolor{JColor}{RGB}{128, 180, 0} 

\IEEEoverridecommandlockouts                              

\title{\LARGE \bf
Words have Weight: Comparing the use of pressure and weight as a metaphor in a User Interface in Virtual Reality
}

\author{Joffrey Guilmet, Suzanne Sorli and Diego Monteiro $^{1}$
\thanks{*Research performed by LII team in EsieaLAB}
\thanks{$^{1}$Corresponding Author}
\thanks{e-mail:{\tt\small diego.vilelamonteiro@esiea.fr}}%
}

\begin{document}

\maketitle
\thispagestyle{empty}
\pagestyle{empty}

\section{Introduction}
Virtual Reality (VR) enables new forms of interaction, often using haptics to convey story- or game-related information. However, haptic feedback can also enhance user interfaces, providing cues not directly tied to the narrative.

While vibration-based feedback is common in controllers and mobile devices, using weight as feedback in interfaces is less explored. Though weight has been studied in VR—especially regarding perception over time and influencing factors—its role in user interfaces needs more attention.

This project investigates how weight can increase the perceived urgency of notifications in VR and whether pressure-based haptics can modulate that perception.

\section{Related Work}

Simulating ungrounded weight in VR is challenging. Thor's Hammer, a hammer-shaped device with internal fans \cite{Heo18}, and Aero-Plane, which uses downward-facing propellers on a controller to generate force \cite{Je19}. Others, like Stroe, use a motor to pull a string connecting the VR controller to the user’s feet \cite{Achberger22}. Other systems~\cite{Monteiro21,Kalus23,Kalus24} opt for liquid mass displacement and positioning, as seen in GravityCup~\cite{Cheng18}, VibroWeight~\cite{Wang22}.

A kind of system that is often ungrounded is the use of Air-Pressure with many examples such as PneuHaptic which uses pneumatically controlled silicone chambers to create varied tactile sensations
\cite{He15} or the work from \cite{Ou16}  which introduces a pipeline for designing and fabricating transforming inflatables using programmable bending mechanisms, enabling interactive applications like wearables and toys. 
However, we can clearly draw inspiration from the works of \cite{Pohl17ACM,Pohl17CHI}, which uses air systems to simulate notifications. Thus, we question: can we use weight and pressure to assist the communication of notifications?
\section{Our System}{

Our haptic system utilizes a closed-loop circuit which are controlled by two stepper-motors actuating two syringes held at 45º , they are managed by Bluetooth through an Unity application. The systems allows refilling with water and air two flexible containers placed in a rigid holder attached to the back of the user's right hand as shown in Figure \ref{fig:system}-a,b. When we activate the syringe, we initially move the water and after  the air is inserted introducing pressure into the same containers, creating a sensation of compression around the hand. This configuration enables us to add weight without pressure and to explore how pressure later affected the perception of weight.

At the hand, the system weighed approximately 119g when empty, and increases to 264g when filled with water. The syringes took 15.15s to fill. The amount of air injected into the system was 100 ml, the measured weight when accounting for air pressure was 350g, at 19ºC 70m altitude.

\begin{figure}[h]
\centering
        \includegraphics[width=1.0\linewidth]{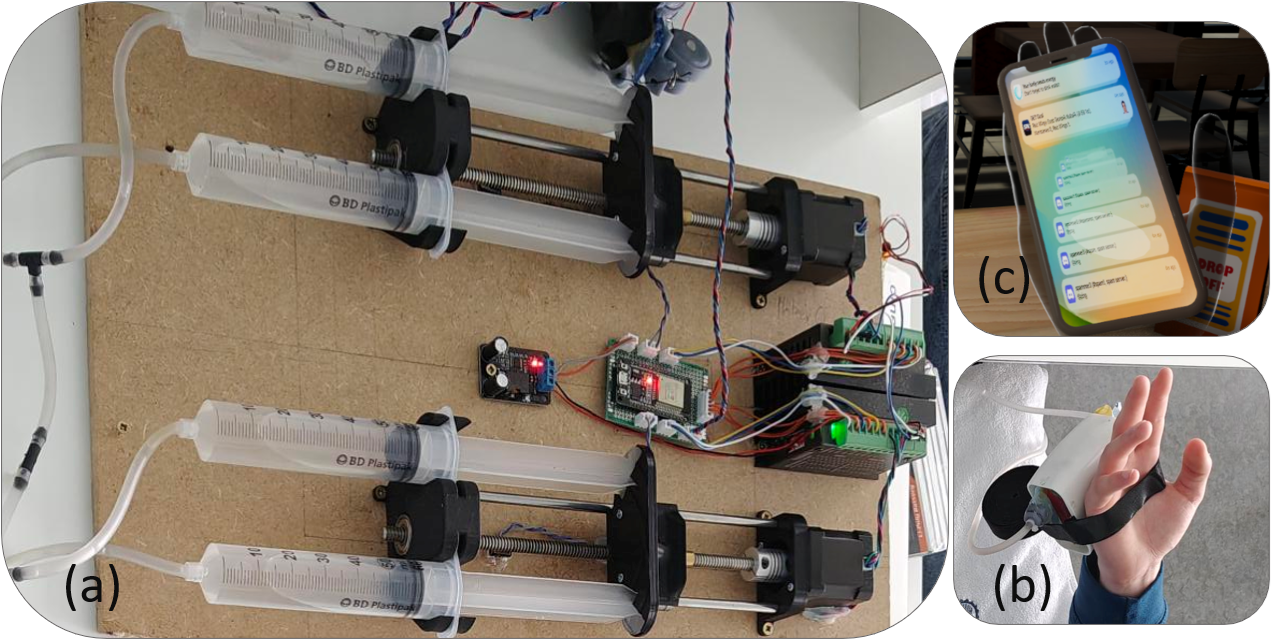}
  \caption{Overview of our system. (a) Stepper-motors are activated to compress syringes that initially contain water, pushing the liquid into two flexible containers placed in a rigid holder attached to the back of the user's right hand. (b) The both containers gradually fills, creating a sensation of increasing weight. Additionally, we introduce air into the same containers to apply pressure on the hand. (c) Simultaneously, the user sees a smartphone in the VR headset displaying a video simulating the reception of a large number of notifications at once.}
\label{fig:system}
\end{figure}
}
\section{Hypothesis}{
 
Given that cutaneous component of weight can be perceived as pressure as well. We hypothesize, that adding a pressure component to a weight transfer system, will increase the perceived weight (H1).

Moreover, we hypothesize that this increased weight will cause participants to feel as if any notifications within a user interface attached to it will be more pressing/urgent (H2).
}
\section{Methodology}
In order to proceed with the project we present an initial ongoing evaluation aimed to evaluate the feasibility of using pressure as a modulator for weight and subsequently as an addition to user interfaces. For this goal we used our system under three conditions: A - Control (No water, or pressure), B - Full Weight and No Pressure , and C - Full Weight and Full Pressure.
 On Unity VR with hand tracking application, the user has a virtual smartphone on their right hand. We played a video on the smartphone synchronized with the filling of the system, in which at every 40 ms, one message and notification would appear.
 \subsection{Measurements}
In order to obtain the subjective feedback we used a series of self-produced questions 1 - "Estimate the value in grams of the object on your hand", and three "In a Scale from 0 to 10, 0 being the lowest" questions 2 - "How coherent was the feeling on your hand and the visual representation", 3 - "How urgent did the messages feel", 4 - "How physically heavy did the messages feel". After, we asked users to rank all versions, based on Weight, Preference and Coherence

\subsection{Procedure}
We recruited 8 participants (2 women) (20 - 32 years old) with an average hand circumference around the palm was of  20.5 cm (SD = 1.24). Except for 1 all had experience with VR. Only 2 had experience with haptics.
In the experiment we had a VR-Ready Computer MSI Katana 15 B13VGK-1685FR Dragon Station and Meta Quest 3S as a HMD, our haptics system, and a measuring tape. The procedure started by obtaining consent and demographic information followed by instructing about the experiment. The participant was then set in position, and the first version was initiated, the experiment order was counter-balanced. The questionnaires were filled when the system was at fully activated, by oral exchange. Upon completion the user was instructed to start the next condition. After all conditions the system was removed and the post-experiment questionnaire was filled, and participants were debriefed.

\section{Results}

A Friedman test (\emph{df}=2) showed significant differences across modes in normalized weight ($\chi^2=14.25, p<0.05$), perceived heaviness ($\chi^2=9.75, p<0.05$), and visual-tactile coherence ($\chi^2=6.05, p<0.05$), but not in perceived urgency ($\chi^2=0.67, p=0.1$). Kendall's $W$ indicated strong agreement on weight perception ($W=0.89$), and good agreement on preference ($W=0.77$) and coherence ($W=0.75$).

Mode A consistently ranked lowest in weight and preference. Mode C was rated heaviest by 7/8 participants and most preferred by 5/8. Mode B fell between A and C in both weight and preference.

Wilcoxon Signed-Rank tests confirmed that Mode A differed significantly from B and C in normalized weight and heaviness (all $p<0.05$). However, no significant difference was found between B and C for heaviness ($p=0.1$). For coherence, Mode C outperformed A ($p<0.05$), but not B ($p=0.1$). Urgency ratings did not differ significantly across any pairs (all $p=0.1$).

\section{Discussion and Future Work}
In this work, we were able to validate H1 but not H2. Our results show that pressure influences the sensation of weight in our context, their specific relationship and limits needs exploration. Future work should test different pressure systems to deepen this understanding.

Many participants rated non-weighted systems as having maximum urgency for notifications, leaving little room for haptic modulation. As a next step, we plan to experiment with varying urgency levels or explore other types of information to better assess the impact of weight-based feedback.

\bibliography{references}




\end{document}